\documentstyle{article}
\title{{\bf The zeropoint field --- no longer a ghost.}}
\author{Trevor~W.~Marshall\\
Dept. of Mathematics, University of Manchester,\\
Manchester M13 9PL, U.K.}
\date{\today}
\begin{document}
\maketitle
\begin{abstract}
We develop a local realist analysis of parametric down
conversion, based on the recognition that the pump field,
instead of down converting spontaneously, does so through
its nonlinear coupling with a real zeropoint,
or ``vacuum" electromagnetic field. 
The theory leads to the prediction of a new
phenomenon --- that, in addition to the main
down-conversion rainbow, there is a satellite
rainbow, whose intensity is about 3 per cent
of the main one. Confirmation of this prediction
will call seriously into question the current
description of the light field in terms of photons.
\end{abstract}
\section{Introduction}
Light is nowadays widely believed to be made of
photons, that is discrete packets of energy $\hbar\omega$,
where $\hbar$ is Planck's constant divided by $2\pi$. This
is ironic, because Max Planck  actually opposed quite
strongly\cite{kuhn} the concept of light quanta from 1905, when
Albert Einstein first proposed them (as a ``heuristic"
hypothesis), and for at least 12 years thereafter. Planck
also proposed, in 1911, the concept of a real zeropoint
electromagnetic field, which he offered  explicitly as
an alternative to the light quanta\cite{milonni,mex}.
Planck's constant enters into this theory in the
role of the scaling of the zeropoint spectrum.
In 1951 Einstein\cite{besso}
himself conceded that there was something very
peculiar about photons, when he said ``Nowadays
every Tom, Dick and Harry thinks he knows what a
photon is, but he is wrong". So, finally, Planck
and Einstein perhaps came close to agreement. The
peculiarities referred to by Einstein have since
become much more acute, to the extent that even
supporters of the photon concept are now using
words like ``mind boggling"\cite{ghz} and ``absurd"\cite{bz}
to describe their properties.

The properties of photons which have merited the
description of ``absurd" originate in a certain strange
behaviour of all quantum particles, first pointed
out in 1935 by Einstein, Podolsky and Rosen\cite{epr}
and by Schr\"{o}dinger\cite{schrod}, and known
nowadays by the general term {\it entanglement}.
According to this, if a pair of quantum
particles is prepared in a {\it superposition state},
then a measurement made on one particle causes
an instantaneous change in the state of the other,
which means that quantum mechanics is in conflict
with Special Relativity. It is curious that the
only real experimental support for this strange
set of events comes from such ephemeral
``particles" as photons; reliable evidence from
such as atoms, electrons and nuclei has steadily
evaded us so far. I shall show that, at least in
the substantial experimental domain of
parametric down conversion (PDC), there
is a rational and consistent theory of
light, based on Planck's concept of the zeropoint field,
in which photons are  entirely absent. Indeed it may be
claimed that we are returning to the unquantized Maxwell
theory of light.
\section{The photon description of PDC}
PDC occurs when a beam of coherent light from a laser,
often referred to as the pump,
is incident on a nonlinear optical crystal. A
rainbow emerges from the crystal, each frequency,
or colour, being emitted in a certain direction.
More specifically, if the pump frequency is
$\omega_0$,
then the PDC rainbow contains all frequencies
less than $\omega_0$, and
the angle  at which a given frequency $\omega$
emerges depends on $\omega$ and $\omega_0$.
For example, if the pump's wavelength is 300nm, which is
in the near ultraviolet,
and the crystal is potassium dihydrogen phosphate (KDP),
then the down converted light
at 450nm, which is blue, will be at an angle of
8.2 degrees to the pump, while that at 600nm (yellow) will
be at 10.5 degrees, and that at 900nm (near infrared)
at 16.6 degrees.

The photon description of PDC is disarmingly
simple. A pump photon  $\hbar\omega_0$
down converts spontaneously into a {\it pair} of
photons, $\hbar\omega_1$ and $\hbar\omega_2$, with $\omega_1+\omega_2=
\omega_0$.
So there are certain pairwise correlations (see Fig.1)
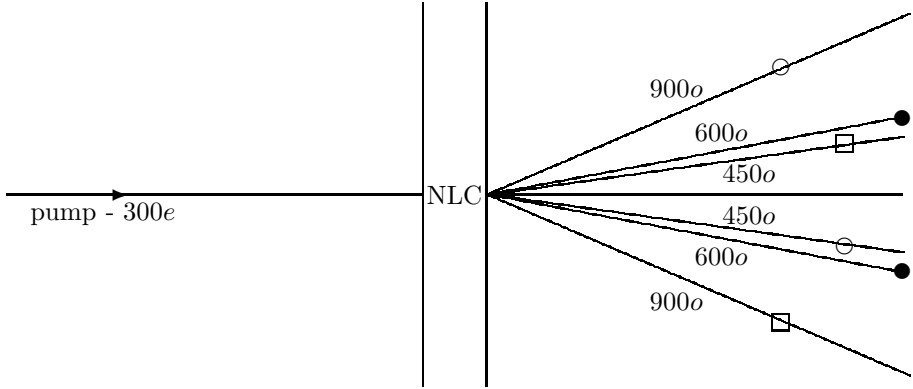
\begin{figure}[htb]
\unitlength 0.85mm
\linethickness{0.4pt}
\begin{picture}(141.67,61.00)
\thicklines
\put(1.00,31.00){\line(1,0){65.00}}
\put(76.00,31.00){\line(1,0){65.00}}
\put(20.0,31.0){\vector(1,0){0.00}}
\thinlines
\put(122.00,51.00){\circle{2.5}}
\put(132.00,23.00){\circle{2.5}}
\put(122.00,11.00){\makebox(0,0){\framebox(2.5,2.5){}}}
\put(132.00,39.00){\makebox(0,0){\framebox(2.5,2.5){}}}
\put(141.00,43.00){\circle*{2.5}}
\put(141.00,19.00){\circle*{2.5}}
\put(66.00,1.00){\line(0,1){60.00}}
\put(76.00,61.00){\line(0,-1){60.00}}
\multiput(76.00,31.00)(0.86,0.12){76}{\line(1,0){0.86}}
\multiput(76.00,31.00)(0.64,0.12){101}{\line(1,0){0.64}}
\put(71.00,31.00){\makebox(0,0)[cc]{NLC}}
\multiput(76.00,31.00)(0.28,0.12){237}{\line(1,0){0.28}}
\multiput(76.00,31.00)(0.86,-0.12){76}{\line(1,0){0.86}}
\multiput(76.00,31.00)(0.64,-0.12){101}{\line(1,0){0.64}}
\multiput(76.00,31.00)(0.28,-0.12){237}{\line(1,0){0.28}}
\put(16.00,28.00){\makebox(0,0)[cc]{pump - 300$e$}}
\put(105.67,14.33){\makebox(0,0)[cc]{900$o$}}
\put(112.67,21.33){\makebox(0,0)[cc]{600$o$}}
\put(117.00,27.67){\makebox(0,0)[cc]{450$o$}}
\put(105.67,47.67){\makebox(0,0)[cc]{900$o$}}
\put(112.67,40.67){\makebox(0,0)[cc]{600$o$}}
\put(117.00,34.33){\makebox(0,0)[cc]{450$o$}}
\end{picture}
\caption{The photon picture of PDC. The suffixes $o$ and $e$ denote
ordinary and extraordinary rays within the crystal (NLC). The two rays
of each pair are denoted by the same symbol.}
\end{figure}
between the rays of down converted light. The angle
$\theta_1$ at which the photon
$\omega_1$ is emitted is determined
through the requirement that the total momentum
carried by the photon pair is equal to the momentum
of the pump photon, that is
\begin{equation}
\hbar\omega_1\sqrt{n_1^2-\sin^2\theta_1}
+\hbar\sqrt{\omega_2^2 n_2^2-\omega_1^2
\sin^2\theta_1}
=\hbar\omega_0 n_0\;,
\end{equation}
or alternatively
\begin{equation}
\omega_1^2\sin^2\theta_1=-\frac{1}{4}n_0^2\omega_0^2
+\frac{1}{2}(n_1^2\omega_1^2+n_2^2\omega_2^2)
-\frac{1}{4n_0^2\omega_0^2}(n_1^2\omega_1^2-n_2^2\omega_2^2)^2\;,
\end{equation}
where $n_i=n(\omega_i)$ is the refractive index of the crystal
at frequency $\omega_i$. The appropriate values of $n_i$,
leading to the values of $\theta_1$ quoted above,
were obtained from \cite{yariv}, Table 16.3. Note that, in
order for this equation to be satisfied,
the pump must be polarized so that it travels through the crystal as an
extraordinary ray, which then gives
$n_0$ less than $n_1$ and $n_2$.
The down converted signals
have been taken to be ordinary rays, which means we
are dealing with {\it Type-I PDC}.

The price paid for the simplicity of this photon
description is the entanglement of the
photon pairs\cite{rarity,zwm,kwiat}, which leads us precisely to the
``mind boggling" and ``absurd" consequences
referred to in the previous section. This price
is too high; indeed Einstein\cite{einst} argued
that it was an abandonment of science!
We shall show in the next section that such an
abandonment becomes unnecessary once we substitute
Planck's zeropoint field for Einstein's photons.
\section{The field description of PDC}
We have shown, in a series of articles\cite{pdc1,pdc2,pdc3,pdc4},
that the above ``photon entanglements" may be consistently
explained as correlations between the wave modes of the
light field. If we accept Einstein's verdict on the
photon ``explanation", we may indeed claim that ours
is the only valid explanation of the whole body of data!
The correlations in question arise because the
corresponding modes of the zeropoint field (see Fig.2)
\begin{figure}[htb]
\unitlength 0.85mm
\linethickness{0.4pt}
\begin{picture}(142.67,61.00)
\thicklines
\put(2.00,31.00){\line(1,0){65.00}}
\put(77.00,31.00){\line(1,0){65.00}}
\put(20.00,31.00){\vector(1,0){0.00}}
\thinlines
\put(122.00,51.00){\circle{2.5}}
\put(132.00,23.00){\circle{2.5}}
\put(122.00,11.00){\makebox(0,0){\framebox(2.5,2.5){}}}
\put(132.00,39.00){\makebox(0,0){\framebox(2.5,2.5){}}}
\put(141.00,43.00){\circle*{2.5}}
\put(141.00,19.00){\circle*{2.5}}
\put(67.00,1.00){\line(0,1){60.00}}
\put(77.00,61.00){\line(0,-1){60.00}}
\multiput(77.00,31.00)(0.86,0.12){76}{\line(1,0){0.86}}
\multiput(77.00,31.00)(0.64,0.12){101}{\line(1,0){0.64}}
\put(72.00,31.00){\makebox(0,0)[cc]{NLC}}
\multiput(77.00,31.00)(0.28,0.12){237}{\line(1,0){0.28}}
\multiput(77.00,31.00)(0.86,-0.12){76}{\line(1,0){0.86}}
\multiput(77.00,31.00)(0.64,-0.12){101}{\line(1,0){0.64}}
\multiput(77.00,31.00)(0.28,-0.12){237}{\line(1,0){0.28}}
\multiput(67.00,31.00)(-0.89,0.12){28}{\line(-1,0){0.89}}
\multiput(26.33,36.67)(-0.89,0.12){28}{\line(-1,0){0.89}}
\multiput(67.00,31.00)(-0.60,0.12){42}{\line(-1,0){0.60}}
\multiput(26.67,38.67)(-0.69,0.12){37}{\line(-1,0){0.69}}
\multiput(42.00,42.67)(0.26,-0.12){98}{\line(1,0){0.26}}
\multiput(67.00,31.00)(-0.11,0.11){3}{\line(-1,0){0.11}}
\multiput(2.33,59.33)(0.29,-0.12){87}{\line(1,0){0.29}}
\multiput(67.00,31.00)(-0.89,-0.12){28}{\line(-1,0){0.89}}
\multiput(26.33,25.33)(-0.89,-0.12){28}{\line(-1,0){0.89}}
\multiput(67.00,31.00)(-0.60,-0.12){42}{\line(-1,0){0.60}}
\multiput(26.67,23.33)(-0.69,-0.12){37}{\line(-1,0){0.69}}
\multiput(42.00,19.33)(0.26,0.12){98}{\line(1,0){0.26}}
\multiput(67.00,31.00)(-0.11,-0.11){3}{\line(-1,0){0.11}}
\multiput(2.33,2.67)(0.29,0.12){87}{\line(1,0){0.29}}
\put(17.00,28.00){\makebox(0,0)[cc]{pump - 300$e$}}
\put(135.00,34.00){\makebox(0,0)[cc]{300$e$}}
\put(135.00,28.00){\makebox(0,0)[cc]{(depleted)}}
\put(106.67,47.67){\makebox(0,0)[cc]{900$o$}}
\put(113.67,40.67){\makebox(0,0)[cc]{600$o$}}
\put(118.00,34.33){\makebox(0,0)[cc]{450$o$}}
\put(106.67,14.33){\makebox(0,0)[cc]{900$o$}}
\put(113.67,21.33){\makebox(0,0)[cc]{600$o$}}
\put(118.00,27.66){\makebox(0,0)[cc]{450$o$}}
\end{picture}
\caption{The field picture of PDC. All modes of
the PDC rainbow are present, but as yet uncorrelated,
in the vacuum. They become correlated  as
a result of the coupling inside the crystal (NLC).
The coupling between modes is as in Fig.1.
The pump mode is depleted, while the others are all enhanced.}
\end{figure}
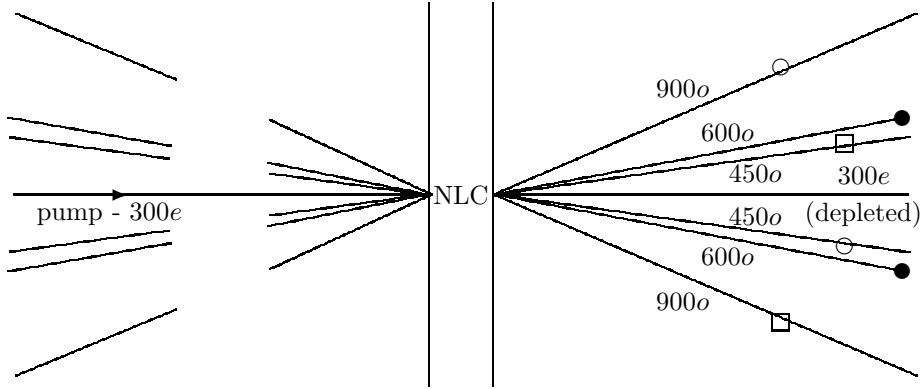
are coupled together inside the pumped crystal.
I remark that the coupling condition of eqn.(1), which
in the photon theory expresses momentum conservation,
becomes a simple classical condition of phase
matching (Ref.\cite{yariv}, Chap.16) on cancellation of $\hbar$.

In Fig.2 we have denoted the zeropoint inputs by interrupted lines.
All ``photon detectors" (and this includes
our own eyes) are blind to these modes; they
register only when
the intensity goes above zeropoint. (This is not
strictly true, because there is always a residual
dark rate, which the photon theory conveniently ignores.
A realist theory of detection, such as ours\cite{pdc4},
must include dark-rate detection events on the same footing
as the detection of ``signals".)
We have shown that the coupling process produces
an intensity enhancement in both the participating
modes, and correspondingly a rather small depletion
in the intensity of the pump, so this explains why
there are counts in the two outgoing channels.

We have established that, within the experimental errors,
the field description of all coincidence experiments,
like Refs.\cite{rarity,zwm,kwiat}, agrees with the photon
description. But, in respect of the much simpler
singles counting rate, there is a prediction made
by the field theory for which the photon theory
has offered no explanation. The rainbow shown schematically in
Figs.1 and 2 {\it does not contain all of the rays actually
emitted by the crystal}, because there is also a less
intense satellite rainbow.

Let us look again at Fig.2 and ask the question ``What
happens if one of the incoming zeropoint modes, for example
the lower $450o$ mode, is replaced by a laser, and the
original laser, that is $300e$, is removed?". Then the
new pump is $450o$,
which produces no PDC rainbow, because the refractive
index of the crystal, for frequencies below that
of the pump, is less than the refractive index
at $450o$, so eqn(1) can not be satisfied.
But the coupling which, in Fig.2,
produced the outgoing pair consisting of the upper
$450o$ and the lower $900o$ should, according to
the field theory, still operate,
because $300e$, like all other modes, is present
in the vacuum. This means that there are
outgoing rays like those depicted in Fig.3. A
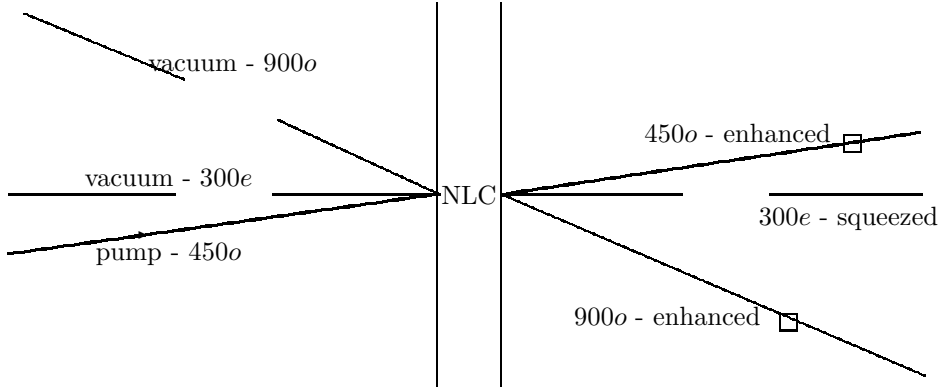
\begin{figure}[htb]
\unitlength 0.85mm
\linethickness{0.4pt}
\begin{picture}(143.00,61.00)
\put(122.00,11.00){\makebox(0,0){\framebox(2.5,2.5){}}}
\put(132.00,39.00){\makebox(0,0){\framebox(2.5,2.5){}}}
\put(67.00,1.00){\line(0,1){60.00}}
\put(77.00,61.00){\line(0,-1){60.00}}
\put(77.00,1.00){\line(0,1){0.00}}
\put(72.00,31.00){\makebox(0,0)[cc]{NLC}}
\multiput(77.00,31.00)(0.28,-0.12){237}{\line(1,0){0.28}}
\multiput(42.00,42.67)(0.26,-0.12){98}{\line(1,0){0.26}}
\multiput(2.33,59.33)(0.29,-0.12){87}{\line(1,0){0.29}}
\put(67.00,31.00){\line(-1,0){25.67}}
\put(26.00,31.00){\line(-1,0){26.00}}
\thicklines
\multiput(67.00,31.00)(-0.86,-0.12){78}{\line(-1,0){0.86}}
\thinlines
\put(77.00,31.00){\line(1,0){28.33}}
\put(119.00,31.00){\line(1,0){24.00}}
\thicklines
\multiput(142.67,40.67)(-0.81,-0.12){81}{\line(-1,0){0.81}}
\thinlines
\put(114.00,40.67){\makebox(0,0)[cc]{$450o$ - enhanced}}
\put(131.33,27.33){\makebox(0,0)[cc]{$300e$ - squeezed}}
\put(103.00,12.00){\makebox(0,0)[cc]{$900o$ - enhanced}}
\put(22.00,24.67){\vector(1,0){0.0}}
\put(25.00,21.67){\makebox(0,0)[cc]{pump - $450o$}}
\put(25.00,33.67){\makebox(0,0)[cc]{vacuum - $300e$}}
\put(35.00,51.67){\makebox(0,0)[cc]{vacuum - $900o$}}
\end{picture}
\caption{Part of the PUC rainbow. A single set of interacting
modes has been selected from Fig.2, and the roles of the input
modes $300e$ and $450o$
have been reversed. As a result of the coupling
the vacuum mode $300e$ is squeezed, while its partner $900o$
is enhanced, producing part of the satellite, or PUC rainbow.}
\end{figure}
detailed calculation\cite{puc1} shows  that,
out of the three outgoing rays, that is
($300e,450o,900o$), the highest-frequency one
is depleted while the other two are enhanced,
which is exactly the same as when $300e$ was the
pump, as in Fig.2. It
may seem rather surprising that the process actually
produces an enhancement, rather than a depletion, of
the pump at $450o$, but it was found that the changes
in the intensities of the $300e$ and $900o$ modes are
only about 3 per cent of those in the main PDC process.
Since, furthermore, these changes are in opposite
directions, the pump enhancement produced
by this process is
extremely small compared with the depletion
in the main process. 

If the pump is
in some polarization intermediate between $e$ and $o$,
then the main PDC rainbow will also be present, and then this
second, weaker one will be observed as a satellite.
As the pump's polarization plane is rotated, the
relative strengths of the main PDC rainbow and
its satellite will vary, the degree of extinction
depending on the way the crystal is cut.
I propose to call this satellite the
{\it Parametric Up Conversion} (PUC) rainbow,
because, although the only {\it detectable} mode
of this rainbow coming out  of the crystal
(in Fig.3 the $900o$ mode) has
a frequency $\omega$ which is less than the pump, its partner
mode (the $300e$ mode) has the frequency
$\omega_0+\omega$. The PUC phenomenon is
well established in classical nonlinear optics (\cite{yariv}, Chap.17),
and my contribution arises from the
recognition that this process must
occur also from the vacuum. Of course, such a weak
signal, at infrared wavelength, is not very easy
to detect, but, subject to the refractive
indices being such that eq.(1) has a solution,
PUC may occur for any frequency less than $\omega_0$.
For example if, for the pump,
we use one of the incoming $600o$ modes of Fig.2,
then an up converted signal at $600o$ will emerge
from the crystal at an angle of 21 degrees to the
pump. This up conversion results from the interaction
of the pump with the two zeropoint modes ($300e,600o$).
\section{Conclusion}
From 1905 to 1917 very few believed in photons, and
it was Planck's view, rather than Einstein's, which
commanded majority support. After  Einstein's famous
article on stimulated and spontaneous emission the
tide began to turn, and after the Compton effect
the majority swung behind Einstein. Maybe the
majority were right; we still have no theory of
the emission of light by atoms which is
consistent with Special Relativity, in the
strict sense required by Einstein, Podolsky and Rosen\cite{epr},
and therefore truly scientific in Einstein's sense\cite{einst}.
The history of science is full of cases
when the majority continued to pay allegiance
to a theory which they knew to be ``absurd"
(to paraphrase Ref.\cite{bz}), so let us hope that
posterity will not judge them too harshly for
making do with an unscientific theory for over 70 years!

Nevertheless, the above explanation of PDC, especially
once its prediction of the new phenomenon has been
confirmed\footnote{I hope to replace {\it once} by {\it now that}
by the time this article is published.}, should encourage
us to try and do better in other areas of optics,
and eventually of all atomic and subatomic physics.

I have reviewed
elsewhere\cite{vig} how, in optics,
nonlocal photon descriptions may be more or
less systematically replaced by local descriptions
which incorporate Planck's field. This systematic
replacement includes, indeed begins with, the
celebrated experiments on Bell-inequality tests
in atomic cascades\cite{marshsant},
which was where ``photon entanglement" was
first observed. Most specialists,
understandably, remained unimpressed, because,
although our treatment of the {\it field} was
unambiguously maxwellian, we  had to improvise,
in a rather crude manner, the description of
the atom-field interaction. Nobody has yet
succeeded in doing, for the Dirac field, what
Planck did for the Maxwell field.

But when it came to PDC we did not need to
know any details about the atom-field interaction;
only the relation between the current and the
field inside the crystal is relevant at optical and
near ultraviolet frequencies. So a purely
maxwellian theory, of the type I have
outlined here,
can engage with the theory depicted in Fig.1,
and which I have called the Photon Theory,
on equal terms.

Our alternative Field Theory is manifestly
local and causal; the incoming fields generate
a current in the crystal, and the outgoing fields
are the retarded fields generated by those currents.

\end{document}